# Task Matters When Scanning Data Visualizations

Laura E. Matzen,[1]* Kristin M. Divis,[1] Deborah A. Cronin[2] & Michael J. Haass[1]

[1]Sandia National Laboratories, [2]University of California, Davis

## ABSTRACT

One of the major challenges for evaluating the effectiveness of data visualizations and visual analytics tools arises from the fact that different users may be using these tools for different tasks. In this paper, we present a simple example of how different tasks lead to different patterns of attention to the same underlying data visualizations. We argue that the general approach used in this experiment could be applied systematically to task and feature taxonomies that have been developed by visualization researchers. Using eye tracking to study the impact of common tasks on how humans attend to common types of visualizations will support a deeper understanding of visualization cognition and the development of more robust methods for evaluating the effectiveness of visualizations.

## 1 INTRODUCTION

What makes a data visualization effective? Evaluating visualizations can be very challenging and is the subject of much research and debate [6,15,20,23,27]. Members of the visualization research community have called for evaluation approaches that assess how well visualizations support their viewers' cognitive needs [5,9,18,30]. From this perspective, an effective visualization successfully exploits its viewers' cognitive processes to draw their attention to relevant information, minimize their attention to irrelevant information, and increase the likelihood of correct interpretation. In order to meet those requirements, visualization designers need to be able to account for the experience, expectations, and biases of the viewer *in addition* to the low-level, perceptual properties of the data visualization.

There is a growing body of research on how the perceptual aspects of visualizations influence viewers' cognitive processes. For example, researchers have demonstrated that increasing the visual saliency of task-relevant information can improve task performance [8,11,12,14,17,19,26] and that changing the visual representation of a dataset can change how viewers interpret it [7] and their biases in interpretation [22]. However, there has been relatively little research on how different high-level tasks impact viewers' attention to different aspects of visualizations. In this paper, we present a simple experiment as an illustration of why this is an important topic in need of additional research.

### 1.1 An Experiment on the Impact of Task

This example is taken from one task in a larger study. In this task, thirty participants recruited from the University of Illinois community were asked to describe either the trend or the outliers in a series of scatterplots. There were 32 scatterplots consisting of four unique plots for each of eight types of trends: positive linear, negative linear, flat, sinusoidal, positive logarithmic, negative logarithmic, positive quadratic, and negative quadratic. The simulated data were drawn from Gaussian distributions and the data points representing the trend were constrained to fall within two vertical standard deviations of the trend function. Half of the plots of each type had two outliers and half had four. The outliers were at least four standard deviations away from the trend function. Examples of the stimuli are shown in Figure 1.

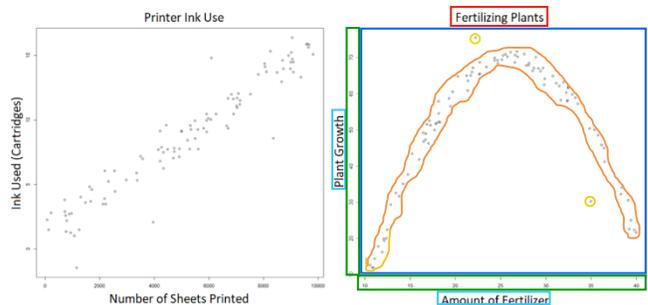

Figure 1: Two representative scatterplot stimuli, one with four outliers (left) and one with two outliers (right). The right image also shows an example of the regions of interest that were used for the eye tracking analysis (Title, Axes, Axis Labels, Trend, Outliers, and Other).

The stimuli were divided into two blocks. For one block, the participants were asked to describe the trend depicted by the scatterplot; for the other block they were asked to describe the outliers. The task-block pairing and the order of the two tasks were counterbalanced across participants. Each scatterplot was shown on a computer screen for 10 seconds, or until the participant pressed a key to advance. While the participants were viewing the scatterplots, their eye movements were recorded with a Smart Eye Pro eye tracker. After the scatterplot disappeared from the screen, the participant verbally described the stimulus from memory.

### 1.2 Results

Two raters independently scored each participant's description of each scatterplot. The participants accurately described the trend for 87% of the trials. When describing the outliers, participants missed one or more of the outliers on 200 out of 398 trials (50.3%). There were only 20 trials in which participants falsely identified an extra outlier (5.0%).

Fixations were calculated using Smart Eye's default algorithm, where any sample for which the velocity over the preceding 200 ms is less than 15°/s is deemed a fixation. The first fixation in each trial was excluded from the analysis, as was any fixation with a duration less than 100 ms. Each stimulus was divided into the following regions of interest (ROIs): Outliers, Trend, Title, Axes, Axis Labels, and Other. The "Other" ROI corresponded to the white space inside of the scatterplot that did not contain any data points. The proportion of fixations to each type of ROI was calculated for each participant and stimulus (Figure 2).

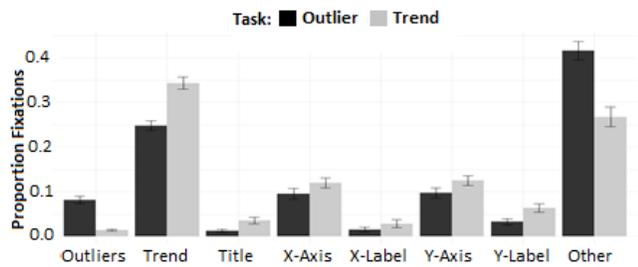

Figure 2: Average proportion of fixations to each region of interest.

The participants' task had a substantial impact on where they allocated their attention within the scatterplots. A mixed effects model (fit with the lme4 package in R software [2]) was used to predict the proportion of fixations as a function of the fixed effects of task and ROI, with random intercepts for subject and stimulus (using Satterthwaite approximation for degrees of freedom). For the trend description task, there were significantly higher proportions of fixations to the Trend ROI as well as the Title, Axis, and Axis Label ROIs. For the outlier description task, the proportion of fixations was significantly higher for the Outlier ROI and the Other ROI (all t-statistics > 2.00 and p-values < .05). The high proportion of fixations to the Other ROI was likely due to participants searching the graphs for outliers as well as the relatively small size of the Outlier ROIs. Similar mixed effects models showed that participants had shorter, more numerous fixations in the outlier task (all $t$s > 10.00, $p$s < .001), which is also indicative of visual search [24]. Figure 3 shows the probability of fixations to each ROI over the time course of the trial. The two tasks produced very different patterns of fixations throughout the trial.

## 2 DISCUSSION

In this task, we observed differences in patterns of eye movements when participants were given different tasks using the same data visualizations. The participants performed the two different tasks successfully, although some outliers were overlooked. The eye tracking data indicated that there were differences in the proportion and timing of fixations to different elements within the graphs. The trend and axes received a relatively high proportion of the participants' fixations, regardless of condition, but the fixations on the outliers was dramatically influenced by the participants' task.

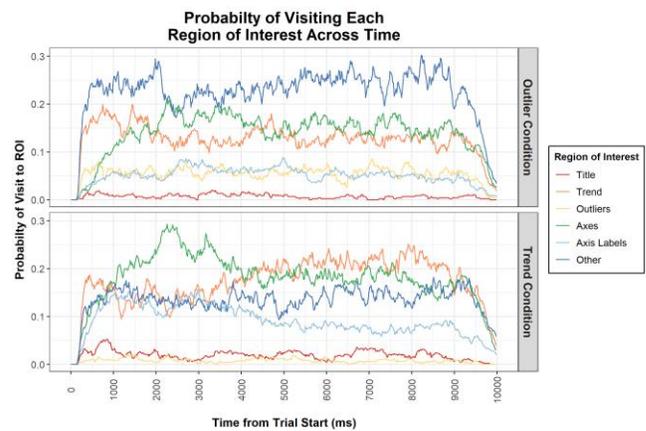

Figure 3. The probability of visiting each ROI over the course of the trial for the two task conditions.

The classic experiment by Yarbus (1967) demonstrated that a person's task and goals impacted their eye movements, and numerous subsequent studies have found similar effects for natural scenes [3,10,13]. Our experiment demonstrates that the same is true for data visualizations. While this is a very straightforward example, it shows how research along these lines could have implications for visualization design and evaluation. In this case, if the designer knew that the outliers might be important to users' tasks, they could choose visual representations that make the outliers more salient and easier to locate. For more complex visualizations, assessing changes in fixation patterns or scan paths for different tasks could reveal patterns that might not be easy to predict. This kind of research would further the understanding of how people make sense of data visualizations and where visual-spatial and cognitive biases [21,28] are most likely to impact their interpretations.

The visualization community has developed numerous taxonomies that break down common visualization types and link them to common tasks cf. [1,4,16,20,25]. These taxonomies could serve as an entry point for visual cognition researchers and as a framework for systematic experimentation. For example, [16] provides a taxonomy of objects and tasks that are common in graph visualization. Researchers could use eye tracking to test how different tasks change patterns of fixations to the same underlying graph objects. This research could support the development of new visualization methods for that domain. It could also support the development of widely applicable evaluation methods that take both bottom-up and top-down features into account to determine whether a graph visualization effectively meets the cognitive needs of its intended users.

## 3 ACKNOWLEDGEMENTS

This work was supported by the Laboratory Directed Research and Development (LDRD) Program at Sandia National Laboratories. Sandia National Laboratories is a multimission laboratory managed and operated by National Technology & Engineering Solutions of Sandia, LLC, a wholly owned subsidiary of Honeywell International Inc., for the U.S. Department of Energy's National Nuclear Security Administration under contract DE-NA0003525. This paper describes objective technical results and analysis. Any subjective views or opinions that might be expressed in the paper do not necessarily represent the views of the U.S. Department of Energy or the United States Government. SAND2020-7549 C.